\documentclass[sigplan,authordraft10pt]{acmart}
\renewcommand\footnotetextcopyrightpermission[1]{}

\AtBeginDocument{%
  }

\usepackage{graphicx} 
\usepackage{lipsum} 
\usepackage{hyperref} 
\usepackage{xcolor}
\usepackage{listings}
\usepackage{todonotes}

\lstdefinestyle{cleanC}{
  language=C,
  frame=single,
  backgroundcolor=\color{gray!5},
  basicstyle=\ttfamily\scriptsize,
  keywordstyle=\bfseries\color{blue!70!black},
  commentstyle=\itshape\color{green!50!black},
  stringstyle=\color{red!60!black},
  numberstyle=\tiny\color{gray},
  numbers=left,
  numbersep=6pt,
  tabsize=2,
  breaklines=true,
  breakatwhitespace=true,
  captionpos=b,
  showspaces=false,
  showstringspaces=false,
  showtabs=false
}

\lstdefinestyle{cleanPython}{
  language=Python,
  frame=single,
  backgroundcolor=\color{gray!5},
  basicstyle=\ttfamily\scriptsize,
  keywordstyle=\bfseries\color{blue!70!black},
  commentstyle=\itshape\color{green!50!black},
  stringstyle=\color{red!60!black},
  numberstyle=\tiny\color{gray},
  numbers=left,
  numbersep=6pt,
  tabsize=4,
  breaklines=true,
  breakatwhitespace=true,
  captionpos=b,
  showspaces=false,
  showstringspaces=false,
  showtabs=false,
  emph={[2]compile},            
  emphstyle={[2]\normalfont\ttfamily\scriptsize}
}

\usepackage{color}  

\usepackage{tikz}
\usetikzlibrary{arrows.meta, tikzmark, calc, positioning, matrix, fit, shapes.misc, decorations.pathreplacing, calligraphy}

\usepackage{comment}

\newcommand{\xtc}{XTC}

\newcommand{\secref}[1]{§~\ref{#1}}
\newcommand{\figref}[1]{Figure~\ref{#1}}
\newcommand{\tabref}[1]{Table~\ref{#1}}

\usepackage{todonotes}
\setuptodonotes{inline}

\title{\xtc, A Research Platform for Optimizing AI Workload Operators}

\author{Hugo Pompougnac}
\affiliation{\institution{Univ. Grenoble Alpes, Inria, CNRS, Grenoble INP, LIG, 38000 Grenoble, France} \country{}}
\email{hugo.pompougnac@inria.fr}

\author{Christophe Guillon}
\affiliation{\institution{Univ. Grenoble Alpes, Inria, CNRS, Grenoble INP, LIG, 38000 Grenoble, France} \country{}}
\email{christophe.guillon@inria.fr}

\author{Sylvain Noiry}
\affiliation{\institution{Univ. Grenoble Alpes, Inria, CNRS, Grenoble INP, LIG, 38000 Grenoble, France} \country{}}
\email{sylvain.noiry@inria.fr}

\author{Alban Dutilleul}
\affiliation{\institution{Univ. Grenoble Alpes, Inria, CNRS, Grenoble INP, LIG, 38000 Grenoble, France} \country{}}
\email{alban.dutilleul@inria.fr}

\author{Guillaume Iooss}
\affiliation{\institution{Univ. Grenoble Alpes, Inria, CNRS, Grenoble INP, LIG, 38000 Grenoble, France} \country{}}
\email{guillaume.iooss@inria.fr}

\author{Fabrice Rastello}
\affiliation{\institution{Univ. Grenoble Alpes, Inria, CNRS, Grenoble INP, LIG, 38000 Grenoble, France} \country{}}
\email{fabrice.rastello@inria.fr}

\begin{document}

\begin{abstract}
  Achieving high efficiency on AI operators demands precise control
  over computation and data movement. However, existing scheduling
  languages are locked into specific compiler ecosystems, preventing
  fair comparison, reuse, and evaluation across frameworks. No unified
  interface currently decouples scheduling specification from code
  generation and measurement. We introduce \xtc{}, a platform that
  unifies scheduling and performance evaluation across compilers. With
  its common API and reproducible measurement framework, \xtc{}
  enables portable experimentation and accelerates research on
  optimization strategies.
\end{abstract}

\maketitle
\pagestyle{plain}

\section{Introduction}
\label{sec:introduction}
For performance engineers and researchers, achieving high efficiency
on AI workloads operators such as matrix multiplication or
convolution is a demanding task. It involves finding a delicate balance
between computation and data movement to ensure that hardware units
are continuously utilized with minimal stalls and idle
time~\cite{tma}.

\subsection{Automation or manual tuning ?}

It is therefore crucial to structure code so that each hardware
resource remains continuously engaged in useful
computation. Typically, the affine loop nests implementing an operator
are transformed through a series of optimizations to enable
vectorization, software pipelining, multi-core multithreading and
other parallelism-based improvements. In addition to exposing
parallelism, transformations must preserve and enhance data locality
by carefully orchestrating data transfers through the cache
hierarchy~\cite{goto}.

Existing approaches span a continuum between fully automated compiler
heuristics and expert-driven manual tuning. Compiler-based
optimizations offer higher productivity but often fail to achieve peak
efficiency when heuristics break down, for instance during
auto-vectorization. Conversely, hand-tuned kernels -- written in
assembly or with intrinsics and delivered as hardware-specific libraries
for compute-intensive tasks, see for instance MKL\cite{mkl} --
reach the highest performance but at the cost of portability,
maintainability, and development effort.  Bridging this gap requires
exposing compiler optimization decisions to experts through interfaces
that are both controllable and portable.

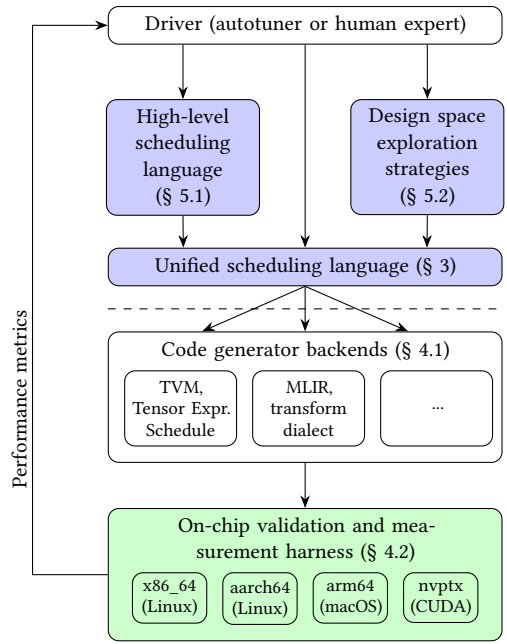
\begin{figure}
    \centering
    \begin{tikzpicture}[
        font=\footnotesize,
        node distance=0.6cm,
        box/.style={rectangle, draw, text width=5cm, text centered, rounded corners, minimum height=0.5cm},
        greenbox/.style={box, fill=green!20},
        bluebox/.style={box, fill=blue!20},
        whitebox/.style={box},
        arrow/.style={draw, -Stealth}
      ]

      \node (driver) [whitebox] {Driver (autotuner or human expert)};

      \matrix (branch) [below=of driver, column sep=12mm] {
        \node[bluebox, text width=1.8cm] (descript) {High-level scheduling language (\secref{sec:decl})}; &
        \node[bluebox, text width=1.8cm] (strategies) {Design space exploration strategies (\secref{sec:strat})}; \\
      };

      \node (transformations) [bluebox, below=0.3cm of branch] {Unified scheduling language (\secref{sec:api})};
      \draw[dashed] ($(transformations.south west) + (-0,-0.3cm)$) -- ($(transformations.south east) + (-0,-0.3cm)$);
      
      \node (generator) [whitebox, below=of transformations] {
        Code generator backends (\secref{sec:backends})\\
        \begin{tikzpicture}[every node/.style={draw, rounded corners, text width=1.3cm,minimum height=1cm, font=\scriptsize}]
          \matrix[column sep=2mm,draw=none, align=center] {
            \node {TVM,\\ Tensor Expr. \\ Schedule}; & \node {MLIR, transform dialect}; & \node {...}; \\
          };
        \end{tikzpicture}
      };
      
      \node (runtime) [greenbox, below=of generator] {
        On-chip validation and measurement harness (\secref{sec:harness}) \\
        \begin{tikzpicture}[every node/.style={draw, rounded corners, text width=0.8cm, font=\scriptsize}]
          \matrix[column sep=2mm,draw=none, align=center] {
            \node {x86\_64 \\ (Linux)}; & \node {aarch64 \\ (Linux)}; & \node {arm64 \\ (macOS)}; & \node {nvptx \\ (CUDA)}; \\
          };
        \end{tikzpicture}
      };

      \draw[arrow] (driver.south -| descript.north) -- (descript.north);
      \draw[arrow] (driver.south -| strategies.north) -- (strategies.north);
      \draw [arrow] (driver) -- (transformations);
      
      \coordinate (tL) at ($(transformations.north) + (-1.62cm,0)$);
      \coordinate (tR) at ($(transformations.north) + ( 1.62cm,0)$);
      \draw[arrow,] (descript.south) -- (tL);
      \draw[arrow] (strategies.south) -- (tR);
      
      \draw [arrow] (transformations) -- (generator);
      \coordinate (gL) at ($(generator.north) + (-1.37cm,0)$);
      \coordinate (gR) at ($(generator.north) + ( 1.37cm,0)$);
      \draw[arrow] (transformations.south) -- (gL);
      \draw[arrow] (transformations.south) -- (gR);
      
      \draw [arrow] (generator) -- (runtime);

      \draw [arrow] (runtime.west) -- ++(-1,0) |- (driver.west) node[pos=0.07,right, align=left, font=\footnotesize,rotate=90] {Performance metrics \\ };

    \end{tikzpicture}
    \caption{\xtc{}'s high-level components and their interactions. \xtc{} allows to decouple research on
      scheduling strategies from code generation, validation and measurement.}
    \label{fig:xtc_arch}
\end{figure}

\subsection{Scheduling languages}

Scheduling languages~\cite{sched_survey} advance this goal by allowing
experts to script optimization transformations. They provide a
programmable interface for triggering transformations such as tiling
or fusion, thereby reducing reliance on opaque compiler
heuristics. These transformations can be manipulated either directly
by an engineer or through autotuning.

Although existing scheduling languages share similar abstractions and
primitives, each remains closely tied to a specific compilation
ecosystem. For example, TVM's Tensor Expression (TE) scheduling
interface~\cite{tvm} and the Halide language~\cite{halide} are tightly
integrated into their respective frameworks. Similarly, MLIR's
Transform dialect~\cite{transform} is an integral component of
MLIR~\cite{mlir}. Being part of an intermediate representation, the
latter was not designed to be exposed to programmers and is
challenging to interact with.

In practice, final performance of generated code depends both
on the richness of exposed transformations and on the quality of
heuristics embedded in opaque compiler passes. In other words, from a
research perspective, it is important to decouple the scheduling
components from the code generation ones.

There is currently no unified, user-facing API flexible enough to
decouple scheduling specification from code generation. Consequently,
adopting one of these languages typically limits access to features
available in other software stacks. Even comparing technologies
against each other is difficult due to the lack of a shared
representation and infrastructure, particularly for obtaining
comparable performance metrics.

\section{Contributions}
\label{sec:contribution}
To address these limitations, we propose \xtc{} -- a research and
prototyping platform designed to decouple scheduling and code
generation, therefore allowing the researcher to experiment at the
level of scheduling strategies. As shown in \figref{fig:xtc_arch}, it
provides three entry points (in blue) to control operator scheduling
while leveraging the code generation infrastructure of mainstream
toolchains. \xtc{} then provides a multi-platform measurement
framework (in green) exposing a range of software and hardware
performance counters.

\textbf{Unified API for scheduling.}
We introduce a unified API that abstracts core components from
multiple scheduling languages, enabling seamless integration of
diverse scheduling compilers. On top of this API, we expose a
lightweight, higher-level declarative scheduling language to
simplify manual experimentation.

\textbf{Unified API for measurement.}
We distribute \xtc{} with a controlled measurement setup that
minimizes variability and exposes detailed hardware performance
metrics. This cross-platform harness -- to our knowledge, the first to
access hardware performance counters on Apple Silicon CPUs in addition
to (i) x86 and non-Apple ARM CPUs, and (ii) NVIDIA GPUs -- ensures
reproducible and quantitative comparisons across compilation
pipelines. Using this infrastructure, we compare results against
optimized C baselines and peak performance bounds, validate
cross-backend consistency by replaying identical schedules through
multiple code generators, evaluate performance models against actual
measurements, and demonstrate end-to-end gains within the
Aidge~\cite{aidge} framework on complete neural networks.

\textbf{Integration with state-of-the-art backends.}
The unified API interfaces with existing scheduling frameworks such as
TVM and MLIR, allowing researchers to leverage the rapidly evolving
infrastructure of large and possibly complex software ecosystems. The
interface also supports user-defined backends, facilitating the
evaluation of research prototypes.

\textbf{Space exploration.}
We provide interfaces for automating design space exploration and
evaluation, enabling experts to connect high-level scheduling
strategies with custom sampling and predictive models. This interface
makes it straightforward to reproduce the search spaces of
state-of-the-art autotuners, such as Ansor\cite{ansor}.

By integrating a unified API, interoperable backends, autotuning
interfaces, and a cross-platform measurement runtime, \xtc{} enables
reproducible performance research and facilitates the comparison of
scheduling strategies across hardware and software stacks.

\section{A unified API for scheduling AI workloads}
\label{sec:api}
This section explains how to schedule an operator with \xtc{}.
We first show how to attach a schedule to an operator, then
present the scheduling primitives exposed by
\xtc{}'s API.

\subsection{Schedules and operators}

\xtc{} provides a fixed set of common AI operators (\texttt{matmul},
\texttt{conv2d}, \texttt{relu}, \texttt{padding}, \texttt{transpose})
which share hyper rectangular and unordered iteration spaces.
They are combined into computation graphs and optimized through classic
loop-nest transformations (strip-mining, splitting, reordering,
etc.). In this framework, given a loop nest that implements a linear
algebra operator, the goal of scheduling is to transform the loops in
order to make more effective use of the target resources.

\begin{figure}
\centering
\begin{lstlisting}[style=cleanC]
for (int I = 0; I < 256; I++)
 for (int J = 0; J < 258; J++)
  for (int K = 0; K < 512; K++)
   C[I][J] += A[I][K] * B[K][J];
\end{lstlisting}
\caption{A non-optimized matrix multiplication in C.}
\label{fig:naive_matmul}
\end{figure}

\begin{figure}
\centering
\begin{lstlisting}[style=cleanC]
for (int I = 0; I < 256; I+=1) {
 for (int J = 0; J < 256;  J+=16) {
  VLOAD(vectC00,C[I][J]);
  VLOAD(vectC01,C[I][J+8]);         
  for (int K = 0; K < 512;  K+=4) {
   VBROADCAST(vectA00,A[I,K]);
   VLOAD(vectB00,B[K,J]); 
   VFMADD(vectC00,vectA00,vectB00);
   VLOAD(vectB01,B[K,J+8]); 
   VFMADD(vectC01,vectA00,vectB01);
   
   VBROADCAST(vectA01,A[I,K+1]);
   VLOAD(vectB00,B[K+1,J]); 
   VFMADD(vectC00,vectA01,vectB00);
   VLOAD(vectB01,B[K+1,J+8]); 
   VFMADD(vectC01,vectA01,vectB01);
   
   VBROADCAST(vectA02,A[I,K+2]);
   VLOAD(vectB00,B[K+2,J]); 
   VFMADD(vectC00,vectA02,vectB00);
   VLOAD(vectB01,B[K+2,J+8]); 
   VFMADD(vectC01,vectA02,vectB01);
   
   VBROADCAST(vectA03,A[I,K+3]);
   VLOAD(vectB00,B[K+3,J]);
   VFMADD(vectC00,vectA03,vectB00);
   VLOAD(vectB01,B[K+3,J+8]);
   VFMADD(vectC01,vectA03,vectB01);
  }
  VSTORE(C[I][J],vectC00);
  VSTORE(C[I][J+8],vectC01);
 }
 for (int J = 256; J < 258; J++)
  for (int K = 0; K < 512; K++)
   C[I][J] += A[I][K] * B[K][J];   
\end{lstlisting}
\caption{
  The target implementation of the matrix multiplication in
  \figref{fig:naive_matmul}. CPU-dependent instructions
  (\textit{i.e.} intrinsics) are hidden behind C macros.
}
\label{fig:target_implem}
\end{figure}


As a running example, consider the matrix multiplication in
\figref{fig:naive_matmul}, for which we want to generate an optimized
implementation as shown in \figref{fig:target_implem} (in C for the
sake of readability). The outer loop $I$ stays unchanged (line~1). We
split $J$'s iteration space at $256$, yielding a main block $[0,256)$
(lines~2--32) and a short remainder $[256,258)$ (lines~33--35). This
ensures the main block has an extent along $J$ that is a multiple of
$16$, enabling vector-friendly register tiles (line~2). With 8-wide
SIMD, a 16-element tile emits two full vectors per inner step. Along
the reduction dimension $K$, we use a register-tile size of $4$
(line~5). We then permute loops so that $K$ precedes $J$, fully unroll
$K$, and vectorize $J$ to execute the inner kernel with SIMD
instructions.

Hand-writing such C implementation -- even with preprocessor helpers
-- is error-prone and brittle, and profiling variants (e.g., without
unrolling along $K$, with tiling along $I$, and so on) requires
revisiting non-trivial code.

\begin{figure}
  \centering
  \begin{lstlisting}[style=cleanPython, literate={{\#}}{{\#}}1]
import xtc.graphs.xtc.op as O
from xtc.backends.mlir import Backend

I, J, K, dtype = 256, 258, 512, "float32"
a = O.tensor((I, K), dtype, name="A")
b = O.tensor((K, J), dtype, name="B")
with O.graph(name="mm_graph") as gb:
  O.mm(a, b, name="mm0")

impl = Backend(gb.graph)
sch = impl.get_scheduler()

sch.dims = ['I','J','K']
sch.split(root="mm0", dim="J", segments={"J[0]":0,"J[1]":256})
sch.strip_mine(root="J[0]",dim="K", tiles={"K1": 4})
sch.strip_mine(root="J[0]",dim="J", tiles={"J1": 16})
sch.unroll(root="J[0]",unrolls={"J1": 16, "K1": 4})
sch.vectorize(root="J[0]",axes=["J1"])
sch.interchange(root="mm0",
      permutation=["I","J[0]","J[1]"])
sch.interchange(root="J[0]",
      permutation=["K","K1","J1"])
sch.interchange(root="J[1]",permutation=["K"])
  
comp = impl.get_compiler()
module = comp.compile(sched)
executor = module.get_executor()
res = executor.execute()
  \end{lstlisting}
  \caption{
    The scheduling of a matrix multiplication using the \xtc{} API (MLIR backend).
  }
  \label{fig:api_python_example}
\end{figure}

In contrast, \figref{fig:api_python_example} expresses the same
transformation sequence with the \xtc{} Python API. The operator
(\texttt{mm}, for \emph{matrix multiplication}) is defined in lines
4--8. Lines 25--28 trigger compilation and execution. Lines~13–23
specify the loop transformations; parameters (tile sizes, loop
reordering, vectorization, etc.) are exposed as Python knobs.
We also introduce a \emph{root} label to disambiguate branches created by
explicit splits (e.g. \texttt{'J[0]'}, \texttt{'J[1]'}). Before any
split, the root is the operator id (\texttt{mm0} here).

This representation of the scheduling state relies on a single
Python/C++ object (the variable \texttt{sch}, defined on
line~11), inspired by TVM schedules.

\begin{figure*}
\centering
\begin{lstlisting}
(*@\textcolor{red}{func.func @myfun(}@*)
 (*@\textcolor{red}{\%A: memref<256x512xf32> \{llvm.noalias\},}@*)
 (*@\textcolor{red}{\%B: memref<512x258xf32> \{llvm.noalias\},}@*)
 (*@\textcolor{red}{\%C: memref<256x259xf32> \{llvm.noalias\}}@*)
(*@\textcolor{red}{)\{}@*)
 (*@\textcolor{blue}{linalg.matmul \{mymm\}}@*)
  (*@\textcolor{blue}{ins(\%A, \%B : memref<256x512xf32>, memref<512x258xf32>)}@*)
  (*@\textcolor{blue}{outs(\%C : memref<256x258xf32>)}@*)
 (*@\textcolor{red}{return}@*)
(*@\textcolor{red}{\}}@*)
tr.named_sequence @__tr_main(%arg0: !tr.op {tr.readonly}) {
 (*@\textcolor{blue}{\%mm}@*) = tr.structured.match attributes {mymm} in %arg0 : (!tr.op) -> !tr.op
 (*@\textcolor{blue}{\%t\_mm}@*),(*@\textcolor{green}{\%loop}@*) = tr.structured.tile_using_for (*@\textcolor{blue}{\%fst}@*) sizes [1,0,0]
 (*@\textcolor{blue}{\%fst}@*), (*@\textcolor{blue}{\%snd}@*) = tr.structured.split (*@\textcolor{blue}{\%t\_mm}@*) after 256  {dimension = 1 : i64} : !tr.op
 (*@\textcolor{blue}{\%t\_mm0}@*),(*@\textcolor{green}{\%loop1}@*),(*@\textcolor{green}{\%loop2}@*) = tr.structured.tile_using_for (*@\textcolor{blue}{\%first}@*) sizes [0,16,4]
  : (!tr.op) -> (!tr.op, !tr.op, !tr.op)
 (*@\textcolor{blue}{\%t\_mm4}@*),(*@\textcolor{green}{\%loop5}@*) = tr.structured.tile_using_for (*@\textcolor{blue}{\%t\_mm0}@*) sizes [0,0,1]
  : (!tr.op) -> (!tr.op, !tr.op)
 (*@\textcolor{blue}{\%t\_mm6}@*),(*@\textcolor{green}{\%loop7}@*),(*@\textcolor{green}{\%loop8}@*) = tr.structured.tile_using_for (*@\textcolor{blue}{\%snd}@*) sizes [0,1,1] : (!tr.op) -> (!tr.op, !tr.op, !tr.op)
 tr.structured.vectorize (*@\textcolor{blue}{\%t\_mm4}@*) : !tr.op
 (*@\textcolor{red}{\%f}@*) = tr.get_parent_op (*@\textcolor{green}{\%loop}@*) {isolated_from_above} : (!tr.op) -> !tr.op
 tr.apply_patterns to (*@\textcolor{red}{\%f}@*) {
  tr.apply_patterns.vector.reduction_to_contract
  tr.apply_patterns.vector.transfer_permutation_patterns
 } : !tr.op
 tr.apply_patterns to (*@\textcolor{red}{\%f}@*) {
  tr.apply_patterns.vector.lower_outerproduct
  tr.apply_patterns.vector.lower_contraction
 } : !tr.op
 tr.loop.unroll (*@\textcolor{green}{\%loop2}@*) {factor = 4 : i64} : !tr.op
 tr.yield 
}
\end{lstlisting}
\caption{
  An example of an MLIR Transform dialect script applied to a matrix
  multiplication equivalent to the one in
  \figref{fig:naive_matmul}. The transformation state is carried by MLIR
  variables. In blue, the evolving state of the \texttt{linalg.matmul} operator
  being optimized. In green, the evolving state of the low-level loops
  (\texttt{scf.for}) materialized during the tiling. In red, the evolving state
  of the function containing the operator (\texttt{func.func @myfun}).
}
\label{fig:transform_example}
\end{figure*}

By comparison, the MLIR Transform dialect operates at a lower
level. The state is distributed across SSA values that are created,
consumed, and sometimes invalidated throughout the transformation
sequence; see \figref{fig:transform_example}. These values may denote
the operator, the loops materialized along the way, or the enclosing
function. Such IR-centric interface also surfaces IR constraints (SSA
discipline, types) and adds boilerplate. By comparison, \xtc{}'s API
makes the schedules easier to express and reason about. In our
framework, Transform scripts are generated when targeting MLIR
rather than written by hand.

\subsection{Scheduling primitives}

\begin{table*}[h!]
  \centering
  \footnotesize
  \begin{tabular}{|c|c|c|}
    \hline
    \xtc{} primitive & TVM/TE counterpart                           & MLIR transform dialect counterpart               \\ \hline
    Strip mine       & \texttt{split}                               & \texttt{tile\_using\_for} (1D)                   \\ \hline
    Interchang       & \texttt{reorder}                             & Implicitly carried by the dataflow of the script \\ \hline
    Unroll           & \texttt{unroll}                              & \texttt{loop.unroll}                             \\ \hline
    Vectorize        & \texttt{vectorize}                           & \texttt{vectorize} + \texttt{apply\_patterns}    \\ \hline
    Parallelize      & \texttt{fuse} + \texttt{parallel}            & \texttt{tile\_using\_forall} (1D)                \\ \hline
    Split            & \textcolor{gray}{\texttt{loop\_partition}}   & \texttt{split\_handle} + \texttt{split}          \\ \hline
    Pack             & \texttt{cache\_read} + \texttt{compute\_at}  & \textcolor{gray}{\texttt{pack}}                  \\ \hline
    Bufferize        & \texttt{cache\_write} + \texttt{compute\_at} & \textcolor{gray}{\texttt{pack}}                  \\ \hline
    Fuse             & \texttt{compute\_at}                         & \textcolor{gray}{\texttt{fuse\_into\_containing\_op}}           \\ \hline
  \end{tabular}
  \caption{The scheduling primitives exposed in XTC and their
    counterparts in TVM/TE and the MLIR Transform dialect. The
    implementations colored in gray are still experimental.}
  \label{fig:api_python}
\end{table*}

As illustrated in \tabref{fig:api_python}, the API of \xtc{} exposes
ten scheduling primitives.  This scheduling interface is an extensible
proposal, designed to adapt to new needs. In particular, achieving
high performance on GPUs will require new primitives that are not yet
implemented (although XTC already supports profiling on GPUs).

Each primitive corresponds to a well-known transformation on loop
nests or memory layouts, enabling the systematic construction of
high-performance kernels.  We describe them below, highlighting their
semantics and their correspondence to the backends primitives we
leverage.

\paragraph{Strip mine}
Strip-mining partitions the iteration domain of an affine loop into
regular blocks of fixed size. Consider a loop with induction variable
$i$ iterating from a lower bound $lb$ to an upper bound $ub$ with step
$n$. Strip-mining with a factor $s$, replaces this loop with two
nested loops: the outer loop, with induction variable $i_0$, iterates
from $lb$ to $ub$ in steps of $n \times s$; the inner loop, with
induction variable $i_1$, iterates from $0$ to $(s - 1)$ with step
$n$. The original induction variable is reconstructed as $i = i0 +
i1$. If $(ub - lb)$ is not an exact multiple of $(n \times s)$, the
last block may contain fewer iterations. Note that \emph{tiling} --
which improves locality and reuse, boosting arithmetic intensity -- is
a generalization of strip-mining: it performs multi-dimensional
strip-mining and groups the resulting inner loops in an arbitrary
order.
dimension).

\paragraph{Interchange}
Loop interchange reorders the loops within a nest while respecting
data-dependence constraints and preserving the association of each
loop with its root. It then changes the iteration order, potentially
improving data locality or enabling vectorization.

\paragraph{Split}
Splitting divides a loop's iteration space into multiple contiguous
regions defined by explicit split points. For a loop iterating over
$[lb, ub)$ with step $n$, and a set of split points ${ s_1,
s_2, \ldots, s_k }$ satisfying $lb < s_1 < \dots < s_k < ub$, the
transformation produces $k+1$ loops, each iterating over one segment:
$[lb, s_1)$, $[s_1, s_2)$, $\dots$, $[s_k, ub)$.  Each loop executes
the same body over its restricted range. Unlike strip-mining,
splitting does not introduce additional nesting.  It is particularly
useful to isolate specific regions of the iteration space in order to
apply specific transformations to them -- for example, to vectorize
sections whose size is a multiple of the SIMD width.

\paragraph{Unroll}
Loop unrolling expands a loop by replicating its body as many times as
there are iterations, handling the induction variable explicitly
within the inlined code.  Unrolling exposes instruction-level
parallelism and reduces loop-control overhead.

\paragraph{Vectorize}
Vectorization maps the computation of a loop or a loop nest onto the
SIMD resources of the target hardware (and implicitly unrolls the
concerned dimensions).  The loops must first be tiled so that each
iteration block matches the SIMD register width. When supported, this
primitive also triggers FMA formation.

\paragraph{Parallelize}
Parallelization distributes loop iterations across multiple threads or
cores. The parallelized loop must correspond to an inherently parallel
dimension and typically appear among the outermost loops. An MLIR
compiler such as \texttt{mlir-opt} targets OpenMP directives, whereas
TVM uses pthreads by default. 

\paragraph{Pack/Bufferize}
\emph{Packing} is applied at a certain loop level, on one of the input
tensors. It allocates a new local memory and copies the elements that
will be used by the operator below the loop, in the order of their
access. This transformation is a trade-off between improving the
spatial locality and the effectiveness of the hardware prefetcher, and
the time and memory spent performing this reordered copy. We allow the
padding of the new local memory, in order to reduce the conflict
misses. \emph{Bufferization} can be seen as a symmetry of packing,
with the same trade-off. A new local buffer is created at the
specified loop level, that will be used to store the newly computed
data of the output tensor. Once the computation of the loop is done,
the data from the local buffer is copied to the output tensor, while
modifying its ordering to fit the original layout.

\paragraph{Fuse}
\emph{Fusion of producer} rematerializes the computation of the
producer of the input buffer. \emph{Fusion of consumer} on the other
side brings computation of the consumer operation.

\paragraph{Implementation limitations}
TVM and MLIR both distinguish between the level of abstract tensors
(TE in TVM, \texttt{tensor} dialect in MLIR) and that of access to
persistent memory (TIR in TVM, \texttt{memref} dialect in MLIR).  In
practical terms, \texttt{pack} and \texttt{fuse\_into\_containing\_op}
are not provided at the \texttt{memref} level but rather at
the \texttt{tensor} level. Similarly, \texttt{loop\_partition} is not
provided at the TE level but at the TIR level. In \xtc{}, support for
TIR on the TVM side, and for linalg on tensors on the MLIR side, is
still experimental and is not exposed at the moment.

\section{Infrastructure}
\label{sec:infra}
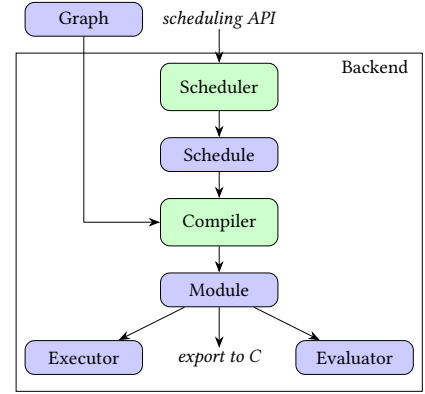
\begin{figure}
    \centering
    \tikzset{
      node distance=1.0cm,
      box/.style={rectangle, draw, text width=1.5cm, text centered, rounded corners, minimum height=0.5cm},
      minibox/.style={rectangle, draw, text width=1.5cm, text centered, rounded corners, minimum height=0.5cm},
      greenbox/.style={box, fill=green!20, minimum height=0.7cm},
      bluebox/.style={minibox, fill=blue!20},
      arrow/.style={draw, -Stealth},
      container/.style={box, inner sep=0.5cm},
    }

\scalebox{0.9} {
\begin{tikzpicture}[node distance=0.8cm, font=\footnotesize]
    \node[greenbox] at (0,0) (scheduler) {Scheduler};
    \node[bluebox] at (0,-1) (schedule) {Schedule};
    \node[greenbox] at (0,-2) (compiler) {Compiler};
    \node[bluebox] at (0,-3) (module) {Module};
    \node[bluebox] at (-2, -4) (executor) {Executor};
    \node[bluebox] at (2, -4) (evaluator) {Evaluator};
    
    \draw (-3,0.5) rectangle (3,-4.5);
    \node at (2.3,0.3) (backend) {Backend};
    
    \node[bluebox] at (-2,1) (graph) {Graph};
    
    \draw[arrow] ([yshift=0.5cm] scheduler.north) -- (scheduler.north);
    \node at (0,1) {\textit{scheduling API}};
    \draw[arrow] (scheduler) -- (schedule);
    \draw[arrow] (schedule) -- (compiler);
    \draw[arrow] (compiler) -- (module);
    \draw[arrow] (module) -- (executor);
    \draw[arrow] (module) -- (evaluator);
    \draw[arrow] (module.south) -- ++(0,-0.6cm) node[midway, below=0.2cm] {\textit{export to C}};
    \draw[arrow] (graph) |- (compiler);
   
\end{tikzpicture}
}
    \caption{\xtc{}'s backend structure. Green boxes represents classes applying transformations, blue boxes are the input or produced objects.}
    \label{fig:xtc_backend}
\end{figure}

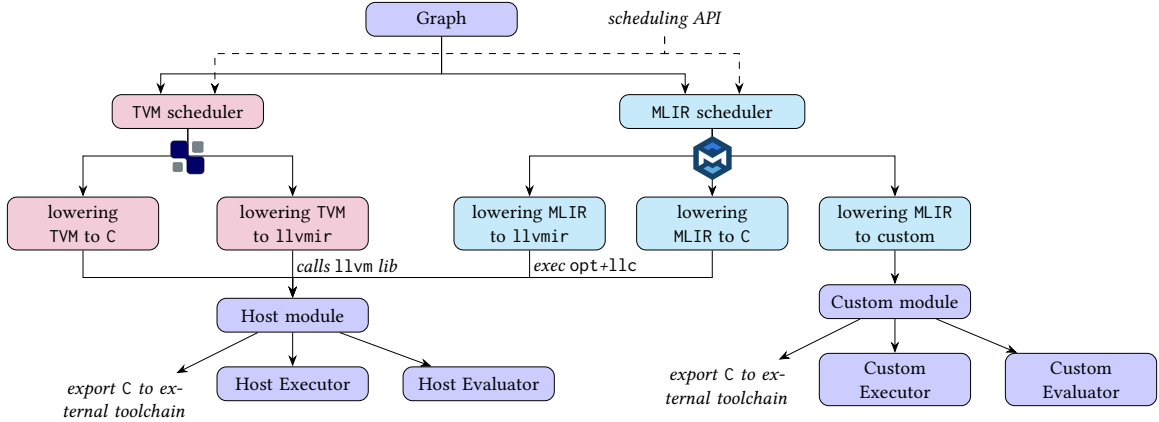
\begin{figure*}
    \definecolor{tvm_color}{RGB}{150 115 166}
    \definecolor{mlir_color}{RGB}{36,67,106}
    \centering
    \tikzset{
      box/.style={rectangle, draw, text width=2.0cm, text centered, rounded corners, minimum height=0.5cm},
      hiddenbox/.style={text width=2.0cm, text centered},
      minibox/.style={rectangle, draw, text width=2.0cm, text centered, rounded corners, minimum height=0.5cm},
      greenbox/.style={minibox, fill=green!20},
      tvmbox/.style={minibox, fill=purple!20},
      bluebox/.style={minibox, fill=blue!20},
      mlirbox/.style={minibox, fill=cyan!20},
      arrow/.style={draw, -Stealth},
      dashedarrow/.style={draw, -Stealth, dashed},
    }
    
\scalebox{0.9}{
    \begin{tikzpicture}[node distance=0.5cm, font=\footnotesize]

    \node[bluebox] (graph) {Graph};
    
    \node[tvmbox, below left=0.85cm and 1.5cm of graph] (tvm_sched) {\texttt{TVM} scheduler};
    \node[mlirbox, text width=2.5cm, below right=0.85cm and 1.5cm of graph] (mlir_sched) {\texttt{MLIR} scheduler};
    
    \node[right=1.2cm of graph.east] (scheduling) {\textit{scheduling API}};
    \draw[dashedarrow] (scheduling.south) |- ([xshift=0.4cm, yshift=0.6cm] tvm_sched.north) -- ([xshift=0.4cm] tvm_sched.north);
    \draw[dashedarrow] (scheduling.south) |- ([xshift=0.4cm, yshift=0.6cm] mlir_sched.north) -- ([xshift=0.4cm] mlir_sched.north);
    
    \draw[arrow] (graph.south) |- ([xshift=-0.4cm, yshift=0.3cm] tvm_sched.north) -- ([xshift=-0.4cm] tvm_sched.north);
    \draw[arrow] (graph.south) |- ([xshift=-0.4cm, yshift=0.3cm] mlir_sched.north) -- ([xshift=-0.4cm] mlir_sched.north);
    
    \node[tvmbox, below left=1.0cm and -0.7cm of tvm_sched] (tvm_c) {lowering \texttt{TVM} to \texttt{C}};
    \node[tvmbox, below right=1.0cm and -0.7cm of tvm_sched] (tvm_llvmir) {lowering \texttt{TVM} to \texttt{llvmir}};
    
    \draw[arrow] (tvm_sched) |- ([yshift=0.6cm] tvm_c.north) -- (tvm_c.north);
    \draw[arrow] (tvm_sched) |- ([yshift=0.6cm] tvm_llvmir.north) -- (tvm_llvmir.north);
    \node[below=-0.2cm of tvm_sched] (tvm_logo) {\includegraphics[width=1.0cm]{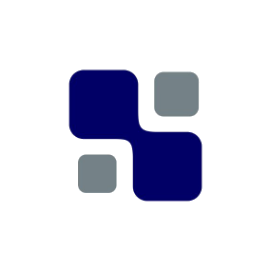}};

    \node[mlirbox, below left=1.0cm and 0.2cm of mlir_sched] (lowering_llvmir) {lowering \texttt{MLIR} to \texttt{llvmir}};
    \node[mlirbox, below=1.0cm of mlir_sched] (lowering_c) {lowering \texttt{MLIR} to \texttt{C}};
    \node[mlirbox, below right=1.0cm and 0.2cm of mlir_sched] (custom_lowering) {lowering \texttt{MLIR} to custom};
    
    \draw[arrow] (mlir_sched) |- ([yshift=0.6cm] lowering_llvmir.north) -- (lowering_llvmir.north);
    \draw[arrow] (mlir_sched) |- ([yshift=0.6cm] lowering_c.north) -- (lowering_c.north);
    \draw[arrow] (mlir_sched) |- ([yshift=0.6cm] custom_lowering.north) -- (custom_lowering.north);
    \node[below=-0.2cm of mlir_sched] (tvm_logo) {\includegraphics[width=1.0cm]{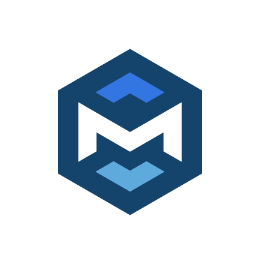}};
    
    \node[bluebox, below=0.7cm of tvm_llvmir] (host_module) {Host module};
    \node[hiddenbox, text width=2.5cm, below left=0.6cm and 0.0cm of host_module] (host_export_c) {\footnotesize \textit{export \texttt{C} to external toolchain}};
    \node[bluebox, below=0.5cm of host_module] (executor) {Host Executor};
    \node[bluebox, below right=0.5cm and 0.5cm of host_module] (evaluator) {Host Evaluator};
    
    \node[bluebox, below=0.5cm of custom_lowering] (custom_module) {Custom module};
    \node[hiddenbox, text width=2.5cm, below left=0.6cm and 0.0cm of custom_module] (custom_export_c) {\footnotesize \textit{export \texttt{C} to external toolchain}};
    \node[bluebox, below=0.5cm of custom_module] (custom_executor) {Custom Executor};
    \node[bluebox, below right=0.5cm and 0.5cm of custom_module] (custom_evaluator) {Custom Evaluator};
    
    \draw[arrow] (tvm_c.south) |- ([yshift=0.3cm] host_module.north) -- (host_module.north);
    \draw[arrow] (tvm_llvmir.south) |- ([yshift=0.3cm] host_module.north) -- (host_module.north);
    \draw[arrow] (lowering_llvmir.south) |- ([yshift=0.3cm] host_module.north) -- (host_module.north);
    \draw[arrow] (lowering_c.south) |- ([yshift=0.3cm] host_module.north) -- (host_module);
    \node[below right=0.0cm and -1.2cm of tvm_llvmir] (llvm_bindings) {\footnotesize  \textit{calls \texttt{llvm} lib}};
    \node[below right=0.0cm and -1.2cm of lowering_llvmir] (opt_llc) {\footnotesize \textit{exec \texttt{opt}+\texttt{llc}}}; 

    \draw[arrow] (custom_lowering) -- (custom_module.north);
   
    \draw[arrow] (host_module) -- (executor);
    \draw[arrow] (host_module) -- (evaluator);
    \draw[arrow] (host_module) -- (host_export_c);

    \draw[arrow] (custom_module) -- (custom_executor);
    \draw[arrow] (custom_module) -- (custom_evaluator);
    \draw[arrow] (custom_module) -- (custom_export_c);

    \end{tikzpicture}
}
    \caption{Implementation of the existing \texttt{TVM} and \texttt{MLIR} backends in \xtc{}. The custom \texttt{MLIR} target shows how an accelerator with an offloading runtime could be added. Blue boxes are the input/produced objects. The \texttt{TVM} boxes are in colored in purple, while the \texttt{MLIR} ones are in cyan.}
    \label{fig:xtc_existing_backends}
\end{figure*}

This section presents the components involved in compiling an
operation graph together with its schedule, the components responsible
for evaluating the generated code, and the interactions between them.

\subsection{\xtc{} backends}
\label{sec:backends}

Our interface currently supports two scheduling pipelines:
\begin{itemize}
\item \emph{TVM schedules on Tensor Expressions (TE)}, where tensors
  are manipulated directly (rather than through TVM-IR, which
  materializes control flow);
\item the \emph{MLIR Transform dialect} applied to operations on
  \texttt{MemRef} values (i.e. explicit persistent memory).
\end{itemize}

We refer to these different pipelines as \emph{backends}. A backend
serves as an abstraction layer for any compilation framework capable
of processing a graph of operations and its associated schedule, each
expressed in the framework's native representation. A backend applies
the schedule and generates an executable for performance evaluation.

In practical terms, both MLIR- and TVM-based pipelines ultimately
generate LLVM IR, which is eventually processed using the LLVM
toolchain (\texttt{opt}, \texttt{llc}). They can alternatively produce
C source code for export. However, \xtc{}, while emitting optimized C
code, does not enforce vectorization and alias information. This
aspect should be improved as it is of interest to provide efficient C
code generation for a number of tools (intrumentation, analysis,
debugging).

From a software architecture perspective, when a \texttt{Graph} is
created, its associated \texttt{Scheduler} records the scheduling API
calls and builds an internal representation of the schedule. It then
applies this schedule and compiles the resulting program into an
executable \texttt{Module}.

Integrating a new backend into \xtc{} requires providing a class that
implements the \texttt{Scheduler} interface and producing an
executable that conforms to \xtc{}'s ABI: a function named after the
graph and taking as parameters the graph's inputs and outputs,
each passed as a contiguous raw pointer.

\figref{fig:xtc_backend} illustrates the general structure of a
backend. Each component is detailed below.

\paragraph{Schedule}
The unified scheduling API is implemented by subclassing the abstract
\texttt{Scheduler} class. Each subclass generates backend-specific
scheduling instructions. For example, calling \texttt{unroll} emits a
\texttt{loop.unroll} Transform operation in the MLIR backend.
The resulting instructions form a \texttt{Schedule} object, which is
later consumed by the compiler to transform the graph.

\paragraph{Compiler}
The \texttt{Compiler} takes as input a \texttt{Graph} and its
associated \texttt{Schedule}. It lowers the graph into the
backend's intermediate representation and applies the schedule.  A
target-specific lowering pipeline then generates the final artifact:
an executable, a shared library, or C source code.

\paragraph{Module}
A \texttt{Module} encapsulates the code produced by the compiler and
exposes the necessary runtime facilities. On the host machine, it
loads the generated shared library and manages argument passing and
result retrieval. \xtc{} also supports alternative runtimes, such as
accelerator offloading.

\subsection{Validation and measurement harness}
\label{sec:harness}

Each \texttt{Module} provides an \texttt{Executor} and an
\texttt{Evaluator}. The \texttt{Executor} validates that the optimized
operator produces results consistent with the reference
implementation. The \texttt{Evaluator} generates input tensors,
executes the compiled code, and collects performance
metrics. Profiling may rely on the system's monotonic clock or on
hardware counters.

\paragraph{Accessing CPU hardware counters.}

The evaluation harness is designed as a portable platform for
measuring hardware performance counters.  This abstraction permits
users to easily focus on analyzing execution characteristics to drive
the optimization process.

To initiate measurement, we use human-readable event names to open the
corresponding hardware events.  On all supported plaforms, the
implementation offers comparable measurement capabilities:
\begin{itemize}
 \item On GNU/Linux, event-to-code translation relies 
 on the \texttt{libpfm4} library, with counter access 
 managed through the \texttt{Perf} interface.
 \item On macOS, we leverage Apple's undocumented 
 \texttt{KPerf} system interface and the \texttt{KPep}
 database for translation.
\end{itemize}

Upon completion of the target execution,
the harness halts the counters counting and
retrieves the final event counts.

\paragraph{Profiling GPU code.}
Performance measurement on GPU relies both on profiling the host and
the accelerator. \xtc{} currently only supports NVIDIA GPUs
accelerators for the MLIR backend. The experimental code generation
for GPU leverages the dialects and associated runtime already existing
in upstream MLIR. Thus, it reuses most of our infrastructure built for
the MLIR backend.  The resulting code has a host entry point and
handles offloading of compute kernels. The \xtc's host runtime has been
augmented with hooks that interact with additional performance
measurement libraries.  From the user point of view, there is a single
interface for performance counters, with GPU-related ones being
prefixed to discriminate them from the host ones.  The current
implementation supports modern Nvidia GPUs having a compute capability
higher or equal to 7.5, and it relies on the Nvidia \texttt{CUpti}
library.

\section{\xtc{} for research}
\label{sec:top}
In this section, we show how \xtc{} can be used as a research
platform. Studying the behavior of a given operator or graph on a
given hardware target, as we do in \secref{sec:api} and
\secref{sec:evaluation}, is the most straightforward use case. Beyond
that, we describe two other possible projects:
the implementation of a high-level declarative scheduling language,
and the implementation of exploration strategies for autotuning.

\subsection{Implementing a new scheduling language}
\label{sec:decl}

\begin{figure}
  \centering
  \begin{lstlisting}[style=cleanPython, literate={{\#}}{{\#}}1]
sch.dims = ['I','J','K']
sch.descript({
 'I': [],
  'J[0:256]': {
   'K': [],
    'K#4': ['unroll'],
     'J#16': ['vectorize'] 
  },
  'J[256:258]': {
    'K': []
  }
})
  \end{lstlisting}
  \caption{The reimplementation of the example in
    \figref{fig:api_python_example} using our declarative, high-level
    language.}
  \label{fig:descript_example}
\end{figure}

The \xtc{} API can serve as a foundation for defining higher-level
scheduling languages. Indeed, current scheduling languages remain
imperative and sequential: users must specify transformations
step-by-step, dealing more with compiler internals rather than with
the operator's logic or hardware behavior. The search space becomes
the sequences of transformations rather than the mapping of
computations onto hardware resources, complicating the exploration
strategies of both experts and autotuning algorithms.

\figref{fig:descript_example} illustrates the declarative language we
defined on top of the \xtc{} API. It is basically a reimplementation
of the schedule presented in \figref{fig:api_python_example}. In this
framework, instead of describing a sequence of transformations, one
directly specifies the structure of the target loop nest. The
low-level transformations required to obtain this loop nest are then
inferred automatically.

The language itself is implemented as a dictionary; in this way, it
can be expressed in Python, but also in MLIR, as a dictionary of
attributes. The keys of the dictionary are used to declare
loops. Given a dimension $D$ and an integer $N$, the declaration
$D\#N$ designates a tile of size $N$ along $D$ (lines 6 and 7). Given
a dimension $D$ and two integers $A$ and $B$, the declaration $D[A:B]$
designates the region of a split of size $(A - B)$ along $D$ (lines 4
and 9). A split declaration carries an inner schedule that describes
the scheduling of the operator below this new root (lines 4--8 and
9--11), which does not need to be explicitly named. A bare dimension
(lines 3, 5, 10) designates the outermost loop along a dimension, if
the latter is not a split. The order of declarations in the dictionary
determines the order of loops in the target code.  Finally, unrolling,
vectorization, or parallelization appear as annotations on these
declarations. In practice, they can be represented as values of the
dictionary (lines 6 and 7).

This higher-level form of notation allows the programmer to think in
terms of \emph{target code} rather than transformations, and thus to
reason more naturally about the mapping between the code and the
machine's hardware resources (the innermost tile being mapped on
vector registers, etc.). This approach notably removes the need to
explicitly maintain the temporary state of the transformations (and
their association with a root), whether that state is distributed
across multiple variables -- as in the Transform dialect -- or carried
by a Python/C++ object, as in our scheduling API or in TVM.

\subsection{Defining scheduling strategies}
\label{sec:strat}

While providing different levels of abstraction for scheduling
particular kernels, there is a need for constructing general
scheduling strategies which apply to any operation and can be used to
build features, performance models and design space exploration.

To this end, \xtc{} provides a base \texttt{Strategy} interface which
allows to define a design space based on some scheduling template. The
design space can be for instance a sparse multidimensional grid from
which one must be able to sample, or get default values inferred from
builtin heuristics possibly dependent on a given optimization
level. The implementation relies on the scheduling primitives shown in
\secref{sec:api} to generate a schedule given a sample.

The provided method \texttt{sample(num:int) -> list[Sample]} can be
used for sampling and prediction. When a candidate is chosen, for
instance by some externally defined acquisition function, code
generation is done through the method
\texttt{generate(sch:Scheduler,sample:Sample)} which sets the
scheduler in the desired state. The method
\texttt{default\_schedule(} \texttt{opt\_level:int)} returns a heuristically
determined default given the target properties.

For instance, inspired by the sketches defined by Ansor~\cite{ansor},
this allows to add a flexible and customizable
strategy \texttt{StrategyPRT} which represent tilings, packing, write
caches and fusion. The strategy is defined as a list of tokens given
as a string which represents an outer-inner view of the scheduled
operation. Each token express some free choices of tiles over
multiple dimensions, buffers or fused computations.

In this terminology, the available tokens are, given $Pdims_{1..p}$ the
list of parallel dimensions and $Rdims_{1..r}$ the list of reduction
dimensions:
\begin{itemize}
\item \texttt{T}: tiling all dimensions;
\item \texttt{P}: tiling all $Pdims$;
\item \texttt{R}: tiling all $Rdims$;
\item \texttt{U}: tiling all dimensions with free order;
\item \texttt{O} tiling with order $Pdims_1$, $Rdims$,
  $Pdims_{2..p}$;
\item \texttt{W}: create optionally a write buffer for the
  output;
\item \texttt{B}: create optionally packed buffers for inputs;
\item \texttt{F} optionally fuse some consumers.
\end{itemize}

One can then instantiate \texttt{StrategyPRT("PPWRPRP")} to
represent a design space equivalent to the Ansor sketches for CPU.

\begin{figure}
  \centering
  \begin{lstlisting}[style=cleanPython]
a, b = O.Tensor((256, 128)), Tensor((128, 1024))
with O.graph('matmul_relu') as ctx:
  m = O.matmul(a, b, name='matmul')
  O.relu(m)
graph = ctx.graph
strategy = StrategyPRT(backend.graph, 'PPWRPRP')
samples = strategy.sample(100)
backend = Backend(graph, default_root='matmul')
compiler = backend.get_compiler()
for sample in samples:
   sch = backend.get_scheduler()
   strategy.generate(sch, sample)
   module = compiler.compile(sch.schedule())
   elapsed = module.get_evaluator().evaluate()
   print(sample, elapsed, 'secs')

# Example scheduling primitives for
# sample = [ 1, 16, 4, 4, 1, 16, 16, 1]
sch.strip_mine('i', {'i1': 64, 'i2': 64, 'i3': 4})
sch.strip_mine('j', {'j1': 64, 'j2': 16, 'j3': 16})
sch.strip_mine('k', {'k1': 16})
sch.interchange([
  'i', 'j',   # P
  'i1', 'j1', # P
  'k',        # R
  'i2', 'j2', # P
  'k1',       # R
  'i3', 'j3', # P
])
sch.buffer_at('j1') # W
sch.parallelize(['i', 'j'])
sch.vectorize(['j3'])
sch.unroll({'k1': 16, 'i3': 4})
  \end{lstlisting}
  \caption{
    Example of samples evaluation for the \texttt{PPWRPRP} strategy.
  }
  \label{fig:xtc_strategies_example}
\end{figure}

\figref{fig:xtc_strategies_example} illustrates how to generate samples
from this strategy when scheduling a matrix multiplication and use a
simple random search for evaluating the performance of the graph. It
also shows the equivalent scheduling primitive sequence for an example
sample.

\section{Evaluation}
\label{sec:evaluation}
We present in this section several use cases of \xtc{} as a research
platform. We demonstrate: (i) its performance relative to
hand-written C code, (ii) its ability to leverage multiple compilation
technologies, (iii) its use for evaluating performance models, and
(iv) its integration within an existing machine learning framework.

\subsection{Evaluating \xtc{} against optimized C code}

We first compare the performance of scheduling an operator with \xtc{}
against a hand-written, parameterized C implementation of matrix
multiplication using the Goto strategy~\cite{goto}. This strategy
tiles the operands across cache levels, based on a fixed, vectorized
inner register. We use a register tile of size $4 \times 32$, while
leaving the outer tile sizes free under divisibility constraints. Both
operands are $1024 \times 1024$ matrices, yielding a search space of
594 schedule instances. We then compare the performance of this C
implementation with the corresponding \xtc{} implementation compiled
through the TVM backend.

\begin{figure}
\centering
\includegraphics[width=1.0\linewidth]{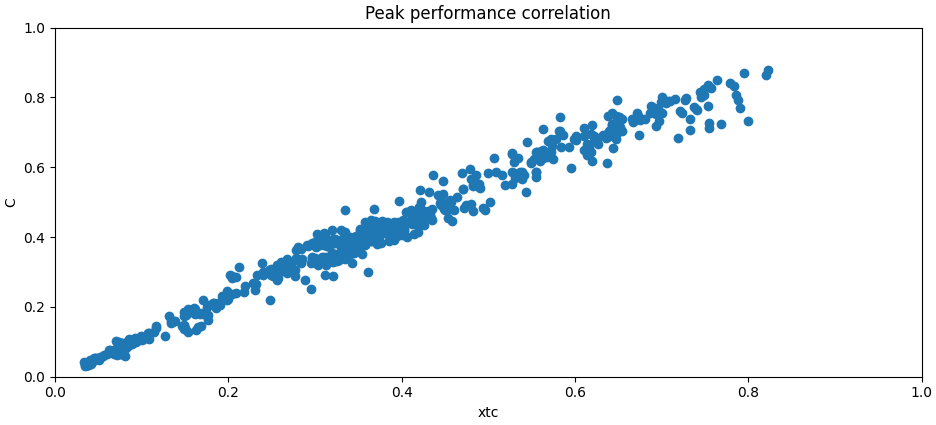}
\caption{\label{fig:xtc_cor_goto}Correlation between \xtc{} with TVM backend and hand optimized C version of the GOTO strategy on a \( [1024, 1024] \times [1024, 1024] \) matrix multiplication.}
\end{figure}

As shown in \figref{fig:xtc_cor_goto}, performance is comparable to
hand-written C code with vector intrinsics. This is not a new result,
though, from a research perspective, it confirms that, at least at the
operator level, using \xtc{} for defining scheduling strategies is a
competitive alternative to hand written C code. In this particular
case, writting and debugging the C code template for this single
strategy took days.

\subsection{Evaluating integration of backends into \xtc{}}

This section discusses how the current integrations of distinct backends into \xtc{} compare against each other. The objective here is not to benchmark \xtc{} against other frameworks, but to exhibit differences between compiler technologies that \xtc{} leverages on.

We consider a single operation and use the previously
described \texttt{StrategyPRT} to generate 100 random schedules. The
strategy is constrained so that the inner tile is always
vectorizable. Each schedule is compiled using both the TVM and MLIR
backends, and we measure the resulting execution times to evaluate
their correlation.

\begin{figure}
\centering
\includegraphics[width=1.0\linewidth]{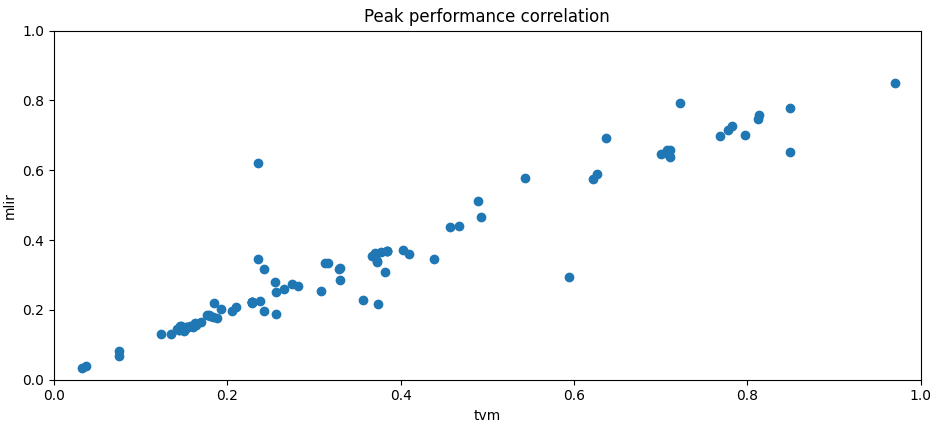}
\caption{\label{fig:xtc_cor_matmul_tu}Correlation for two backends on a \( [512, 128] \times [128, 1024] \) matrix multiplication with \texttt{TU} strategy and vector constraint}
\end{figure}

As shown in \figref{fig:xtc_cor_matmul_tu}, when scheduling a matrix
multiplication with the \texttt{TU} strategy, we get highly correlated
execution times over the explored space. The few cases where relative
performance significantly differs are due to varying unrolling in the
backends.

\begin{figure}
\centering
\includegraphics[width=1.0\linewidth]{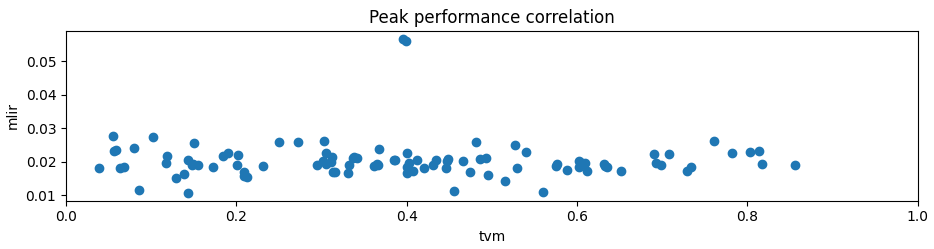}
\caption{\label{fig:xtc_cor_conv_pprprp}Correlation for two backends on a \( [112, 112, 16] \times [7, 7, 3]\) step $2$ conv2d with \texttt{PPRPRP} strategy and vector constraint}
\end{figure}

Conversely, we conducted an experiment consisting of scheduling a
convolution using the \text{PPRPRP} strategy. The result is shown in
\figref{fig:xtc_cor_conv_pprprp}. In this case, the MLIR backend lags
far behind the TVM backend across all samples. This is due to
limitations in the vectorization pass of \texttt{mlir-opt}, which
appears to be disabled when access functions are non-trivial. We were
therefore able to identify this issue (and apply a pre-pass to
simplify the access functions). Of course, other MLIR-based compilers
(such as IREE~\cite{iree} or xDSL\cite{xDSL}) might behave
differently, but this is a typical use case of \xtc{} where one may
seek to identify the limitations of a backend, and optionally improve
it.

\subsection{Evaluating a cache model with \xtc{} }

This section demonstrates how \xtc{} can be used to evaluate
performance models through its search strategies and integrated
hardware counter instrumentation.

\begin{figure}
\centering
\includegraphics[width=1.0\linewidth]{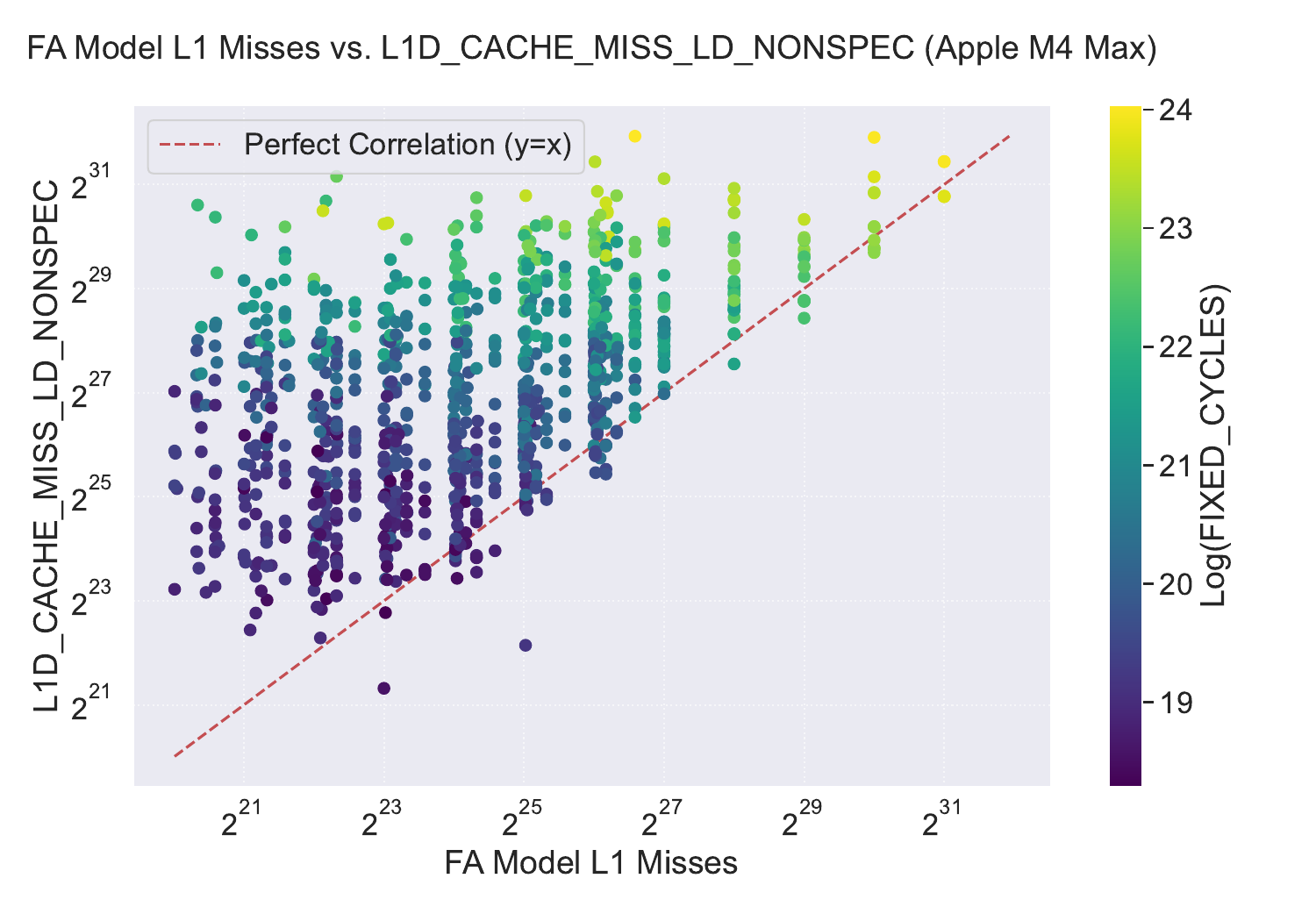}
\caption{\label{fig:xtc_cor_l1}Correlation between L1 cache misses predicted by a full-associative cache model and the corresponding hardware counter on \( [1024, 1024] \times [1024, 1024] \) matrix multiplication samples generated by \xtc{} with the Goto~\cite{goto} strategy running on the Apple M4 Max.}
\end{figure}

Our case study is a fully associative cache model inspired by the one
IOOPT\cite{ioopt} leverages as its cost function. Using \xtc{}, we
assess the model's ability to predict L1 cache misses for different
operator implementations on an Apple M4 Max. \figref{fig:xtc_cor_l1}
and \autoref{tab:correlation_l1_statistics} show the correlation
between predicted and measured L1 cache misses on a range of schedule
instances for a matrix multiplication.

The model exhibits moderate correlation with measured data, capturing
high-miss schedules more accurately than efficient ones.  However, its
overall optimistic prediction struggles to classify across efficient
schedules and to model finer-grained hardware effects.

\begin{table}[h!]
    \centering
    \caption{Correlation statistics for the cache model}
    \label{tab:correlation_l1_statistics}
    \begin{tabular}{lc}
    \toprule
    \textbf{Correlation Type} & \textbf{Statistic ($r$ or $\rho$)} \\
    \midrule
    Pearson ($r$) & $0.534$ \\
    Spearman ($\rho$) & $0.492$ \\
    \bottomrule
    \end{tabular}
\end{table}

\subsection{Evaluating \xtc{} within a full graph export}

Finally, we demonstrate how \xtc{} can be integrated within a complete
inference pipeline. As a proof of concept, we connected \xtc{} to the
Aidge~\cite{aidge} framework, enabling mixed generation of C++
templates and compiled neural network subgraphs. In this setup, \xtc{}
compiles selected subgraphs -- identified as candidates for
optimization -- while the remaining parts are generated through
Aidge's standard C++ flow. Thanks to this approach, one can benefit
from a compiler backend without requiring the full support for all ML
operations. In this particular experiment, we generated inference code
for common convolutional networks (Squeezenet~\cite{squeezenet},
Resnet18~\cite{resnet}, and Resnet50) using \xtc{} with the TVM
backend to compile subgraphs with \texttt{pad},
\texttt{convolution} and \texttt{dense} operators. We show results
for two different CPU architectures, an Intel Core-i5-1135G7
($4.1$GHz) and an ARM RPI4 Cortex-A72 ($1.5$GHz) for single core
inference.

\begin{figure}
\centering
\includegraphics[width=1.0\linewidth]{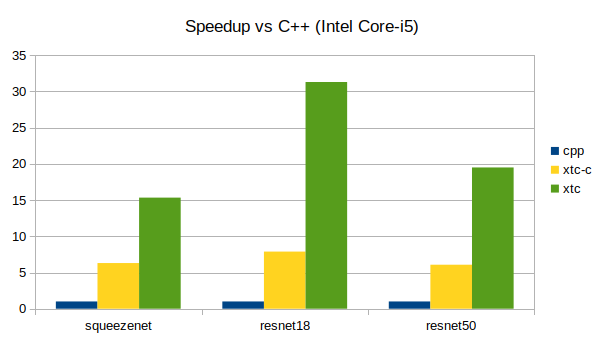}
{\footnotesize (a) On Intel CPU}
\includegraphics[width=1.0\linewidth]{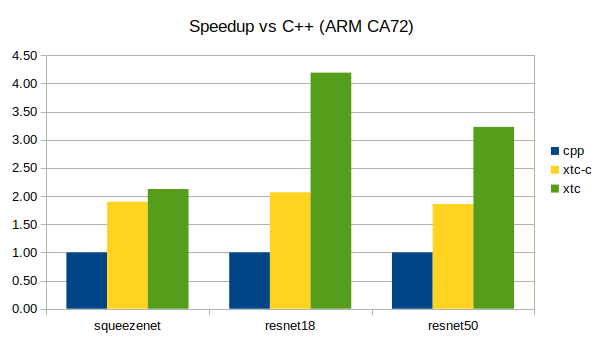}
{\footnotesize (b) On ARM CPU}

\caption{Speedup of partial compilation with \xtc{} against generic C++ export (cpp), in emit C mode (xtc-c) and fully compiled mode (xtc)}
\label{fig:xtc_c}
\end{figure}

In \figref{fig:xtc_c}, we see that using \xtc{} for mapping subgraphs achieve between $\times$15 and $\times$30 speedup on the Intel machine (resp. between $\times$2 and $\times$4 speedup on the ARM machine).

Obviously, there is a large potential for improvements achievable with optimized libraries or ML graph compiler compared to C++ code. With the \xtc{} setup, one can experiment optimization of operations in context and extract subgraphs for fusion strategies.

\section{Related works}
\label{sec:previous}
This section does not attempt to provide an exhaustive survey of
autotuning strategies based on scheduling languages; such coverage is
available in~\cite{sched_survey}.  Instead, we focus on the design
rationale behind \xtc{} and its positioning relative to existing
mainstream infrastructures.

A family of scheduling frameworks exposes purely functional
primitives. Lift~\cite{lift} and Elevate~\cite{elevate} rely on rewrite
rules and strategy combinators; MDH~\cite{mdh} uses algebraic
homomorphisms to express high-level decompositions. XTC intentionally
operates at a lower level. Its primitives denote concrete loop and
memory transformations that correspond to those exposed by
TVM.

TVM \cite{tvm} builds on the distinction between algorithm
specification and scheduling strategy introduced by
Halide \cite{halide}, and generalizes this approach across a wide
range of hardware targets. It is a tensor compiler renowned for
producing high-quality code for many kinds of tensor computations. In
addition, TVM clearly separates graph-level and operator-level
optimizations, enabling end-to-end model execution while supporting
fine-grained operator tuning.

Implementing optimization schemes not natively supported by TVM's
Tensor Expression (TE) language -- such as GotoBLAS-style~\cite{goto}
packing or split transformations~\cite{ttile} -- is non-trivial. TVM's
scheduling language enables experts to define high-level search spaces
via templates, specifying tiling depths, tile sizes, and other
structural parameters. AutoTVM~\cite{autotvm} then orchestrates
sampling, lowering, execution, and modeling to identify performant
configurations within a combinatorial search space.  A key limitation
of AutoTVM is its reliance on dense hyper-rectangular search spaces,
which necessitates \emph{ad-hoc} encodings of sparse constraints
(e.g. divisibility). Despite this, TVM remains a preferred platform
for research due to its code quality and accessible development model.

AutoTVM's requirement for per-operator templates scales poorly to
workloads comprising hundreds of operator variants. This limitation
led to the development of Ansor~\cite{ansor}, which shifts the
template definition to the target architecture rather than the
operator-architecture pair. While this improves scalability, it
couples search strategies tightly to the search space through
hard-coded hyperparameters, making the infrastructure less
extensible. Ansor's emphasis on generality also encourages the use of
scheduling languages that mirror underlying loop transformations. This
design choice eliminates the need to translate abstract search points
into concrete transformation plans, enhancing extensibility and
portability. In theory, adding a new transformation directly extends
the search space.  This approach also facilitates formal reasoning, as
proofs can be constructed at the granularity of individual
transformations. Consequently, frameworks such as Exo~\cite{exo,
exo2}, OptiTrust~\cite{optitrust}, and
Polymorphous~\cite{polymorphous} adopt this one-to-one mapping between
scheduling language constructs and transformation operations.
However, this design introduces control into the search space, thereby
complicating optimization. Moreover, the abstraction is not directly
aligned with the runtime behavior, making metric derivation and
interpretation more opaque.  This hinders the expression of expert
knowledge -- such as inter-operator transfer tuning -- that could
otherwise help structure or prune the search space for faster
convergence.


\xtc{} adopts the philosophy proposed by Mary Hall et al.
in~\cite{sched_survey}, which calls, among other things, for an
effort to (i) unify scheduling languages, and (ii) raise the level of
abstraction of scheduling languages. The scheduling API we propose
aims to address these two challenges by providing unified, general
scheduling primitives to serve as the bedrock for higher-level
representations (e.g. describing the loop structure of the target
code), and for search strategies that are more easily adaptable to
backend-specific constraints.

\section{Conclusion}
\label{sec:conclusion}
We introduced \xtc{}, a research platform for experimenting with
scheduling and performance optimization across compiler frameworks. By
decoupling scheduling from code generation, \xtc{} enables fair
comparison, reproducible measurement, and rapid prototyping of
optimization strategies. Our results show that XTC matches hand-tuned
performance, reveals backend limitations, and allows performance models
evaluation. Beyond its immediate utility, \xtc{} promotes a modular
vision of compiler research where scheduling abstractions, autotuners,
and evaluation tools interoperate seamlessly to accelerate innovation.

\clearpage
\bibliographystyle{IEEEtran}
\bibliography{paper}

\end{document}